\documentclass[preprint]{elsarticle} % <- for arXiv

\usepackage{graphicx}
\usepackage{amssymb}
\usepackage{lineno}

\usepackage{amsmath}
\usepackage{color}
\usepackage{ulem}

\def\klpn{$K^0_L \rightarrow \pi^0 \nu \bar{\nu} \ $}
\def\klpnn{$K^0_L \rightarrow \pi^0 \nu \bar{\nu}$}

\def\klpp{$K^0_L \rightarrow \pi^0 \pi^0  \ $}

\def\klppn{$K^0_L \rightarrow \pi^0 \pi^0$}

\journal{Nuclear Instruments and Methods in Physics Research Section A}

\begin{document}
%\rightline{June 4, 2013}

\begin{frontmatter}

%% Title, authors and addresses

%% use the tnoteref command within \title for footnotes;
%% use the tnotetext command for the associated footnote;
%% use the fnref command within \author or \address for footnotes;
%% use the fntext command for the associated footnote;
%% use the corref command within \author for corresponding author footnotes;
%% use the cortext command for the associated footnote;
%% use the ead command for the email address,
%% and the form \ead[url] for the home page:
%%
%% \title{Title\tnoteref{label1}}
%% \tnotetext[label1]{}
%% \author{Name\corref{cor1}\fnref{label2}}
%% \ead{email address}
%% \ead[url]{home page}
%% \fntext[label2]{}
%% \cortext[cor1]{}
%% \address{Address\fnref{label3}}
%% \fntext[label3]{}

\title{Photon-Veto Counters  
at the Outer Edge of the Endcap Calorimeter for the KOTO Experiment}

%% use optional labels to link authors explicitly to addresses:
%% \author[label1,label2]{<author name>}
%% \address[label1]{<address>}
%% \address[label2]{<address>}

\author[NDA]{T.~Matsumura\corref{ctr1}}
\ead{toru@nda.ac.jp}
\cortext[ctr1]{Corresponding author. Tel.: +81-46-841-3810; fax: +81-46-844-5912.}
\author[NDA]{T.~Shinkawa}
\author[NDA]{H.~Yokota}
\author[Osaka]{E.~Iwai\fnref{preKEK}}
\fntext[preKEK]{Present address: High Energy Accelerator Research Organization (KEK), Japan.}
\author[KEK]{T.K.~Komatsubara}
\author[Osaka]{J.W.~Lee}
\author[KEK]{G.Y.~Lim}
\author[Chicago]{J.~Ma}
\author[Kyoto]{T.~Masuda\fnref{preOka}}
\fntext[preOka]{Present address: Research Core for Extreme Quantum World, Okayama University, Japan.}
\author[Kyoto]{H.~Nanjo}
\author[KEK]{T.~Nomura}
\author[Saga]{Y.~Odani}
\author[Osaka]{Y.D.~Ri}
\author[KEK]{K.~Shiomi}
\author[Osaka]{Y.~Sugiyama}
\author[Saga]{S.~Suzuki}
\author[Osaka]{M.~Togawa}
\author[Chicago]{Y.~Wah}
\author[KEK]{H.~Watanabe}
\author[Osaka]{T.~Yamanaka}

\address[NDA]{\it Department of Applied Physics, National Defense Academy,Yokosuka, Kanagawa~239-8686, Japan}
\address[Osaka]{\it Department of Physics, Osaka University, Toyonaka, Osaka~560-0043, Japan}
\address[KEK]{\it High Energy Accelerator Research Organization (KEK), Tsukuba, Ibaraki 305-0801, Japan}
\address[Chicago]{\it Enrico Fermi Institute, University of Chicago, Chicago, Illinois ~60637, USA}
\address[Kyoto]{\it Department of Physics, Kyoto University, Kyoto~606-8502, Japan}
\address[Saga]{\it Department of Physics, Saga University, Saga~840-8502, Japan}

\begin{abstract}
The Outer-Edge Veto (OEV) counter subsystem for extra-photon detection 
from the backgrounds for the \klpn decay is located 
at the outer edge of the endcap CsI calorimeter of the KOTO experiment at J-PARC.  
The subsystem is composed of 44 counters with different cross-sectional shapes.  
All counters are made of lead and scintillator plates and read out through wavelength-shifting fibers.
In this paper, we discuss the design and performances of the OEV counters 
under heavy load ($\sim$8~tons/m$^2$) in vacuum. 
For 1-MeV energy deposit, the average light yield and time resolution 
are 20.9~photo-electrons and 1.5~ns, respectively. 
Although no pronounced peak by minimum-ionizing particles is observed in the energy distributions,  
an energy calibration method with cosmic rays 
works well in monitoring the gain stability with an accuracy of a few percent. 
\end{abstract}

\begin{keyword}
%% keywords here, in the form: keyword \sep keyword
Lead--scintillator sandwich \sep Energy calibration \sep WLS-fiber readout \sep Kaon rare decay \sep KOTO \sep J-PARC
%% MSC codes here, in the form: \MSC code \sep code
%% or \MSC[2008] code \sep code (2000 is the default)
\end{keyword}

\end{frontmatter}

%%
%% Start line numbering here if you want
%%
%\linenumbers

%% main text

\section{Introduction}

The KOTO experiment is dedicated to observe the \klpn decay 
using the 30-GeV proton beam at Japan Proton Accelerator Research Complex (J-PARC)~\cite{J-PARCacc}.
The \klpn decay is a direct CP-violating and flavour-changing neutral current process.
The branching ratio (BR) is proportional to 
the square of the CP-violation parameter $\eta$ in the CKM matrix~\cite{Walfen} 
and is predicted to be 2.43$\times 10^{-11}$ in the Standard Model (SM)~\cite{Branch}. 
The most attractive feature of this decay is 
the exceptionally small theoretical uncertainty of the BR of only 2--3\%. 
Therefore measurement of the BR %branching ratio 
of this decay mode is highly sensitive to 
%the CP-violating parameter $\eta$ in the SM and 
the contribution of new physics beyond the SM.
The experimental upper limit on the branching ratio was set to be $2.6 \times 10^{-8}$ 
at 90\% confidence level by the E391a experiment at KEK~\cite{E391a}. 
The KOTO experiment~\cite{KOTOpro,Status2,Status,KOTOexp} 
aims to reach a sensitivity below $10^{-11}$ by utilizing the high intensity beam at J-PARC 
and with additional E391a detector upgrades  
to improve the acceptance and the efficiency for background rejection.
%The detector construction was finished and 
%the physics data taking started in May 2013~\cite{Status2,Status,KOTOexp}.  

\begin{figure*}
\begin{center}
\includegraphics[width=15cm]{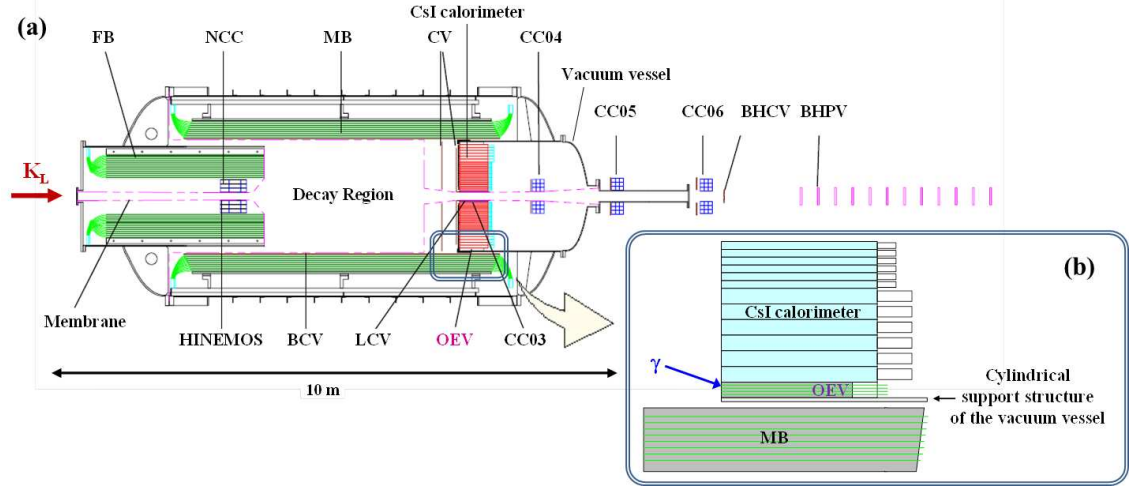}
\caption{
(a) Schematic side view of the detector setup of the J-PARC KOTO experiment. 
Most of the detectors are contained in the cylindrical vacuum vessel of 3.8~m in diameter and 8.7~m in length. 
The $K^0_L$ beam passes through the center of the setup from the left. 
The CsI calorimeter is located downstream of the decay region. 
FB, NCC, MB, OEV, CC03, CC04, CC05, CC06, and BHPV are photon veto counters; 
HINEMOS, BCV, CV, LCV, and BHCV are charged-particle veto counters. 
A membrane separates the decay region, which is in vacuum at $10^{-5}$~Pa, 
from the detector region which is kept at about 0.1~Pa.
(b) Detail of the outer-edge region of the CsI calorimeter. 
The main role of the OEV counters is to detect possible photons 
coming from the \klpp background as indicated by the blue arrow. (Color on-line)
}
\label{fig:KOTOexp}
\end{center}
\end{figure*}

In the KOTO experiment, the \klpnn\ decay is identified 
by detecting two photons from the  $\pi^0$ decay with an electromagnetic calorimeter, 
while the hermetic veto system ensures that no other particles are present. 
The endcap CsI calorimeter located downstream of the decay region (see Fig.~\ref{fig:KOTOexp}a)
serves the function of detecting the two photons. 
All the remaining detectors are photon veto and charged-particle veto counters. 
The well-collimated $K^0_L$ beam at J-PARC~\cite{KOTObeam}
allows us to reconstruct the $\pi^0$ momentum and decay position  
from the energies and positions of the two photons measured with the CsI calorimeter.   
The main background source is expected to be the \klpp decay
(${\rm BR}=8.64\times 10^{-4}$~\cite{Klp0p0}). 
Because the decay can mimic a signal candidate 
if two of the four photons from the $\pi^0\pi^0$ decay are missing,  
%due to detection inefficiency of the photon veto counters, therefore 
the performance of the photon veto counters and the CsI calorimeter are crucial. 

The Outer-Edge Veto (OEV) counter subsystem 
is one of the photon-veto-counter subsystems and 
is located around the outer edge of the endcap CsI calorimeter.  
The 2716 undoped CsI crystals are stacked 
in a cylindrical support structure of the vacuum vessel. 
The OEV counters fill the narrow space 
between the CsI crystals and the cylindrical support structure. 
The main role of the OEV counters is to reject photons 
passing through the outer-edge region of the CsI calorimeter 
before entering the inactive material of the 
support structure (see Fig.~\ref{fig:KOTOexp}b). 
In particular, photons from the \klppn\ decay with an energy around 600~MeV must be detected with high efficiency. 
In fact the kinematics allows one of the two photons from the $\pi^0$ 
to hit the barrel detector (MB in Fig.~\ref{fig:KOTOexp}a) with energy about 10~MeV. 
However this low energy photon may not be detected with a 20\% probability 
due to sampling fluctuations of MB~\cite{KOTOpro}.
To keep a short veto time window under high rate condition, 
the time resolution of a few nanoseconds is required for the OEV counters.
In addition, they need to operate stably in vacuum under the heavy load of the CsI crystals.  

In consideration of the requirements mentioned above, 
we adopted a technology based on lead--scintillator sandwich calorimetry 
with wavelength-shifting (WLS) fiber readout.
This type of detector is efficient for photons 
with energy higher than 100~MeV~\cite{Ineff1,Ineff2}.  
Moreover, it is characterized by a fast response due to the short decay times of plastic scintillator and WLS fiber.  

We describe the design of the OEV counters (Section~\ref{sec:2})
and the construction process (Section~\ref{sec:3}) in detail. 
The following topics on the required performance are discussed in Section~\ref{sec:4}: 
mechanical robustness of the counters located under the CsI crystals,  
discharge characteristic of photomultiplier tubes (PMTs) in vacuum conditions,  
light yield, and time resolution.   
Finally, we discuss an energy calibration method with cosmic rays 
as well as its performance and validity (Section~\ref{sec:5}).

\section{Design}
\label{sec:2}
% of Outer-Edge Veto counters}

\begin{figure}
\begin{center}
\includegraphics[width=8.5cm]{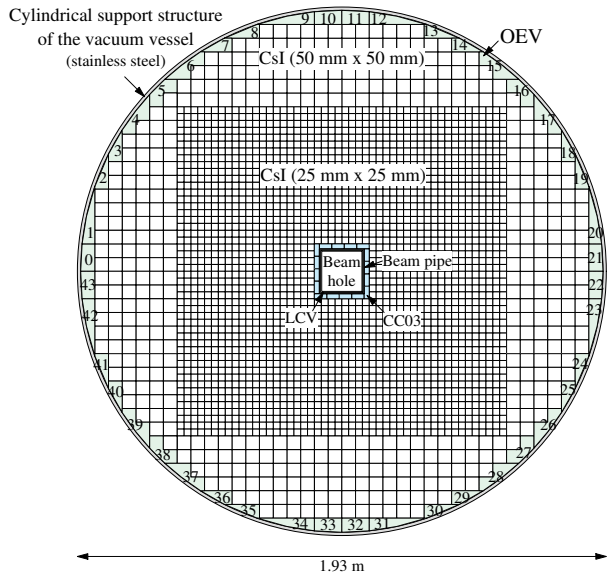}
\caption{Cross section of the endcap CsI calorimeter (view from beam upstream). 
Detector subsystems are contained in the cylindrical support structure 
of the vacuum vessel made of stainless steel. 
The beam pipe is made of carbon-fiber-reinforced plastic (CFRP). 
CC03 and LCV are the inner photon-veto and charged-particle-veto counters.
The inner region of the calorimeter consists of 2240 undoped CsI crystals 
with a 25~mm $\times$ 25~mm cross section, while the outer region consists of 
476 undoped CsI crystals with a 50~mm $\times$ 50~mm cross section. 
OEV counters, which consist of 44 counters with different cross-sectional shapes, 
fill the space between the outer side of the CsI crystals and the vacuum vessel. 
The numbers written on each OEV counter (0--43) are the ID numbers.
}
\label{fig:Endcap}
\end{center}
\end{figure}

Figure~\ref{fig:Endcap} illustrates the upstream view of the endcap of the KOTO detector. 
%Three objects are discussed including 
The CsI calorimeter, the cylindrical support structure of the vacuum vessel, and the OEV counters are shown.
The CsI calorimeter consists of two types of undoped CsI crystals: 
2240 small (25~mm $\times$ 25~mm) and 476 large (50~mm $\times$ 50~mm) crystal blocks.
All crystals are 500~mm long. 
Hamamatsu PMTs (R5330 and R5364) with Cockcroft-Walton (CW) bases~\cite{HVdist} 
are used as readout devices.
The CsI crystals are installed in the cylindrical support structure of the vacuum vessel,   
which is made of 12-mm-thick stainless steel with an inner diameter of 1.93~m. 
OEV is a group of 44 lead-scintillator-sandwich counters. 
They have different cross-sectional shapes  
in order to fill the space between the CsI crystals and the cylindrical support structure. 

The material spares of the barrel detectors FB and MB, 
which were originally built and used in the E391a detector~\cite{MB},  
were utilized for the OEV counters since they shared the same components.   
The first component consists of 5-mm-thick extruded plastic scintillator sheets 
developed for the long barrel detectors. 
This scintillator is based on MS resin (polystyrene 80\%, polymethyl-methacrylate 20\%)~\cite{MS} and 
is suitable to be used under heavy weight load of the CsI crystals.  
The next element consists of 1.5-mm-thick lead sheets which were used for the FB construction.  
We studied the effect of the lead thickness  on detection efficiency
with the constraint of the same counter thickness with a simulation based on Geant4~\cite{Geant4}. 
The results suggested that there was no difference in the rejection power to the \klpp background 
if the lead thickness is between 1~mm and 2~mm. 
We decided to use the 1.5-mm-thick lead sheets for the OEV counters. 
The sampling ratio was estimated to be 23\%. 
The WLS fibers Kuraray Y11(200)M~\cite{Y11} are used to collect scintillation light. 
These 1-mm-diameter fibers were the same as the ones used in MB. 
Readout with the WLS fibers has the advantage of avoiding light-yield non-uniformity originating from  
the short attenuation length (45~cm) of the extrusion-molding scintillator~\cite{MS}.

\begin{figure}
\begin{center}
\includegraphics[width=8.5cm]{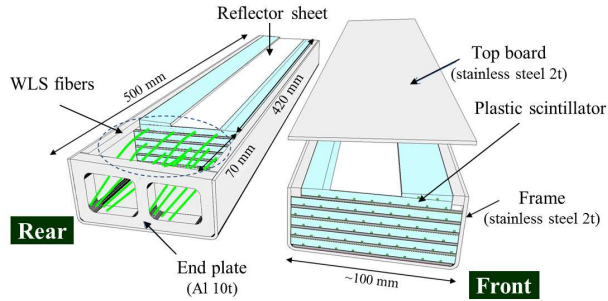}
\caption{Exploded views of an OEV counter (OEV-32), which is located 
at the bottom center of the endcap (see Fig.~\ref{fig:Endcap}). 
The left figure shows the downstream view with the top board and the top reflector sheet removed,  
while the right figure shows the upstream view.  
The area in white color shows the reflector sheet on the second scintillator plate. 
Inside the frame, scintillator plates and lead sheets are stacked alternately.
Light from the scintillator plates is read out through WLS fibers. 
The fibers are extracted through windows in the end plate. (Color on-line)}
\label{fig:module}
\end{center}
\end{figure}

Figure~\ref{fig:module} shows the exploded views of a typical OEV module located at the bottom of the endcap.  
Both the plastic scintillators and the lead sheets are 420~mm long.   
They are stacked alternately in the stainless steel frame. 
The lengths of the frame and the CsI crystals are the same. 
The WLS fibers are bent in the remaining space of the frame 
as a bundle located outside the end plate.
The reflector sheets, Toray Lumirror E60L with reflectivity of 97\%, are 188~$\mu$m thick~\cite{Lumi}.  
They are inserted in direct contact with the whole surface of the scintillator 
to achieve high light-collection efficiency.
Note that we selected this particular reflector 
%because of its stable reflection properties under the load pressure and 
based on our past experience of long-term stable operation with large pressure (15~tons/m$^2$)
in the E391a experiment~\cite{MB}.

The frame structures of the OEV modules at the lower and upper half of the endcap are different. 
The OEV modules placed at the bottom half of the endcap have a robust frame made of 2-mm-thick stainless steel 
(see Fig.~\ref{fig:module}) because they need to support the CsI weight. 
By fastening the stacked layers in the frame tightly,  
warped scintillator plates and/or deformed lead sheets were flattened. 
In contrast, there is no such load on the OEV modules located in the upper half of the endcap. 
Hence, these modules are just covered by 1-mm-thick aluminum plates and bound with polyester tape 
to protect the contents. 
The impact on the detection efficiency of the thin iron frame has been studied with the Geant4 simulation, 
and shown to be negligible for the \klpp background. 
%With the Geant4 simulation, we confirmed that the variation in the detection efficiency ($\varepsilon $)   
%due to the presence of the frame  
%was negligible for photon with energy of 600~MeV,  
%and the efficiency of the veto counter subsystems 
%was estimated to be $1-\varepsilon=1\times10^{-6}$ for this energy region. 

\begin{figure}
\begin{center}
\includegraphics[width=8.5cm]{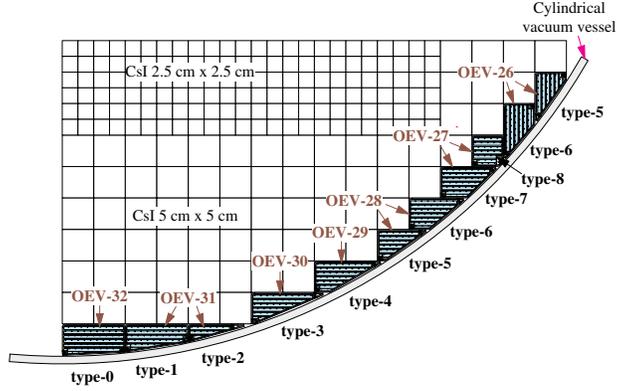}
\caption{
OEV counters located at the bottom right section of the endcap. 
In this figure, each of four counters, OEV-26, 27, 28 and 31
consists of two modules which are read out with a PMT.  
The stacking-layer direction is horizontal at the bottom and vertical at the side.}
\label{fig:bottom_right}
\end{center}
\end{figure}

Hamamatsu R1924A 1" PMTs~\cite{Hamamatsu} were used as readout device. 
A total of 44 PMTs were placed about 80~cm behind the end plate of the counters. 
As shown in Fig.~\ref{fig:bottom_right}, some OEV counters which consist of two modules are connected to one PMT. 
%There are a total of 64 modules with nine different cross-sectional shapes. 
Thus, the 44 OEV counters actually consist of 64 modules with nine different cross-sectional shapes. 

The CW base,  Hamamatsu C10344MOD13, was selected as the high-voltage power supply for the PMTs 
which could operate in the atmospheric pressure or below 0.1~Pa. 
The maximum power consumption of the base is 30~mW, 
which is an order-of-magnitude less than that of the resistor-type dividers. 
This reduces the amount of heat generated inside the vacuum vessel.
Another advantage is that this CW base can be operated 
with the same voltage distributor developed for the CsI calorimeter. 

As illustrated in Fig.~\ref{fig:bottom_right}, 
the direction of the stacked layer is horizontal for the OEV modules 
located at the top and the bottom region  
while the direction is vertical for the side modules.   
%while the direction of the stacked layer is vertical for the side modules,  
%the direction is horizontal for the OEV modules located at the top and the bottom region. 
This serves to reduce the azimuthal angle dependence of the effective radiation length 
for photons emitted from the beam region.

\section{Construction}
\label{sec:3}

\begin{figure}
\begin{center}
\includegraphics[width=8.5cm]{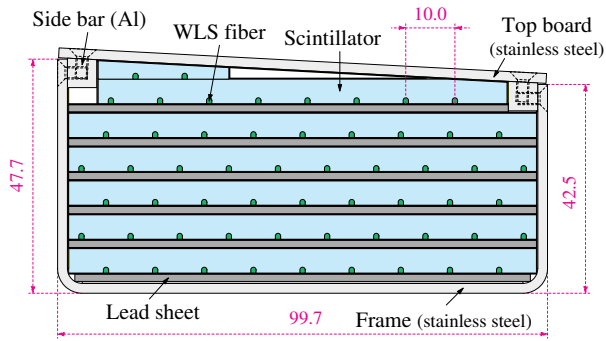}
\caption{Cross section of an OEV module (OEV-33) for the bottom center of the endcap. 
Scintillator plates and lead sheets are stacked in the frame. 
Green circles in the scintillator plates represent the end of the WLS fibers. 
The top plate is fastened to the aluminum side bars % of the frame 
in order to press stacked layers in the frame. 
The module was installed upside down, so that the slope of the top board 
matches the inside slope of the cylindrical vessel. (Color on-line)}
\label{fig:TYPE0}
\end{center}
\end{figure}

Scintillator plates were cut out from a large scintillator sheet with an area of 5.5~m $\times$ 0.68~m. 
They were then shaped to the desired cross sections (rectangle or trapezoid, as shown in Fig.~\ref{fig:TYPE0}) 
with a milling machine and polished with abrasive cloth for optical quality.   
Grooves to embed the WLS fibers were then machined 
on one side of each scintillator plate with an interval of 10~mm. 
The width and the depth of the grooves were 1.2~mm and 1.5~mm, respectively. 
Note that we set the depth somewhat deeper than the WLS-fiber diameter in order to prevent glue from overflowing. 

Ultraviolet curing adhesive, NORAND NOA61~\cite{Norland}, was chosen  
to glue the WLS fibers to the grooves with good optical contact.
The short 30-minute curing time of the adhesive was well suited to mass production. 
We developed an automatic gluing system equipped with 
an X-Y movable stage, the Sigma Koki SGSP46, 
and with a glue dispenser, the Nordson EFD Ultra-1400. 
The control was implemented with a laptop PC via GP-IB communication bus.
We applied 0.25~ml of adhesive to each groove in 10 seconds uniformly without a break. 
Since the viscosity of the adhesive strongly depended on the temperature and elapsed time from its production date, 
making fine adjustments to the air pressure of the dispenser was 
important to maintain the same discharge rate. 
Visible bubbles were removed with air blowers 
after the 145-cm-long WLS fibers were set in the grooves. 
Finally, we irradiated the glue with ultraviolet light ($\lambda=365$~nm) 
with an intensity of 1.5--5.5~mW/cm$^2$. 
It took 30 minutes to deposit the required energy for full curing, 3~J/cm$^2$. 
After extra fibers were cut at the front end, 
the surface was optically polished with a diamond file and an abrasive cloth.

For the bottom modules, special attention was paid to stack the lead--scintillator layers   
to reduce extra space inside the OEV modules. 
A detailed configuration of the layers is illustrated in Fig.~\ref{fig:TYPE0}. 
The scintillator plate with WLS fibers was sandwiched between two sheets of lead in contact with a reflector sheet.  
Prior to the stacking, the thicknesses of all scintillator plates and lead sheets were measured  
with a micrometer with an accuracy of better than 10~$\mu$m. 
Special spacers, such as papers (90~$\mu$m) and/or Lumirror sheets (188~$\mu$m), were layered 
so that the total thickness of all the layers stacked inside the frame was fitted to the designed value within 100~$\mu$m.
The average thickness of the scintillators and the lead sheets of both the top and bottom OEV modules are   
($4.91\pm0.07$)~mm and ($1.44\pm0.03$)~mm, respectively\footnote{
The thicknesses of 82\% of the scintillators and 100\% of the lead sheets 
were measured, being taken into account in the average calculation.}. 
These measured values were specified in the detector simulation code as the material and geometry constants. 
Details of the Monte Carlo simulation code will be described later.

Both the front and rear edges of the scintillators were 
painted with reflector, ELJEN EJ-510~\cite{Eljen}.
The effect of the reflector at the front end 
was tested with a type-8 OEV module by measuring 
the light yield for cosmic rays passing through the center of the module. 
Three different reflectors were tested:  
aluminized Mylar (direct attachment), 
Lumirror E60L (direct attachment), and EJ-510 (painted). 
The light yields relative to the case without a reflector  
were $1.08\pm0.01$ for the aluminized mylar,  
$1.24\pm0.01$ for the Lumirror sheet, and $1.26\pm0.03$ for EJ-510.
Although there was no significant advantage of EJ-510 over Lumirror on the light yield, 
we selected the painted reflector because of the ease of handling.% (no cutting and taping)

\begin{figure}
\begin{center}
\includegraphics[width=8.5cm]{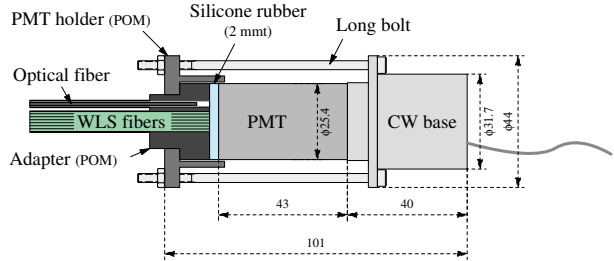}
\caption{Illustration of the WLS-fiber bundle and PMT (Hamamatsu R1924A) 
with CW base (C10344MOD13) for an OEV counter.   
The WLS-fiber bundle is glued to the adapter made of POM. 
By fastening a PMT holder made of POM to the PMT, 
the end of the WLS-fiber bundle is tightly contacted to the PMT window 
through a 2-mm-thick silicone rubber (EJ-560). 
Sleeve of optical fiber is fastened to the adapter of the WLS fiber and optical fiber is hold in the sleeve. }
\label{fig:PMT}
\end{center}
\end{figure}

As shown in Fig.~\ref{fig:PMT}, the fiber bundle\footnote{
Teflon sealing tape was wrapped around the fiber bundle for protection,  
and the tape was covered with black tape for light shield.}  
was optically connected to a PMT with a 2-mm-thick transparent silicone rubber, 
ELJEN EJ-560~\cite{Eljen}. 
The end of each WLS-fiber bundle was then glued to an adapter with optical cement, 
ELJEN EJ-500~\cite{Eljen}, and polished with a diamond file and an abrasive cloth.  
Figure~\ref{fig:PMT} also shows the mechanism to fix the fiber bundle to the PMT  
by fastening force through the PMT holder and the adapter\footnote{For the case in which 
two modules were read with a PMT, two semi-cylindrical adapters were used for two bundles.}. 
The adapter and holder were made of black polyoxymethylene (POM) resin 
for its %with considerations of 
mechanical strength and ease of machining. 
To monitor the PMT gain, 
blue light was flashed continuously at 5~Hz through 
an optical fiber of the adapter.  

The PMT holder was fixed to the cylindrical support structure. 
A copper tape (25~mm wide and 0.035~mm thick) was placed around each CW base.  
It was attached to the cylindrical support structure for heat release.
A simple calculation shows that the temperature difference 
is expected to be less than 2 degrees 
by assuming that the length and the heat conductivity of the copper tape 
are 2~cm and 400~W/(m$\cdot$K), respectively.
% base thickness = 35x10^-6 m, tape width = 0.025 m -> area = 8.8e-7 m2
% tape length = 0.02 m
% conductivity = 400 W/(m K)
%The PMT holder was fixed to the cylindrical support of the endcap; 

\section{Performance}
\label{sec:4}
%%%%%%%%%%%%%%%%%%%%%%%%%%%%%%%%%%%%
%%%%%%%%%%%% LOAD TEST %%%%%%%%%%%%%
%%%%%%%%%%%%%%%%%%%%%%%%%%%%%%%%%%%%

\subsection{Load test}

\begin{figure}
\begin{center}
\includegraphics[width=8.7cm]{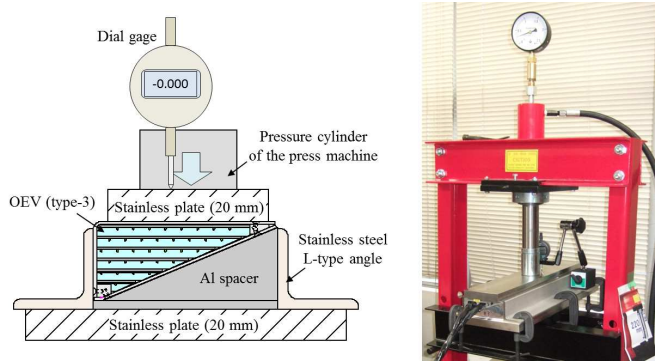}
\caption{
Illustration of the setup of the load test for an OEV module (type-3) and its photograph.
The OEV module placed on the aluminum spacer was pushed 
by the press machine (20~tons at the maximum load) through the top stainless plate, 
and the displacement of the plate was measured by the dial gage. 
The L-type angles, which were clamped to the bottom stainless plate, 
were used to prevent a transverse slide. 
}
\label{fig:loadtest}
\end{center}
\end{figure}

For smooth sliding of the CsI blocks during the CsI-stacking work,  
the level difference of neighboring blocks has to be less than 200~$\mu$m. 
For this reason the structure of the OEV modules located at the bottom part 
should be robust enough to support the weight of the column of CsI crystals above them  
%The maximum load is 406~kg weight ($\sim$8~tons/m$^2$) 
without causing large downward deformation 
(the load by a single tower of the crystals reaches 406-kg weight, equivalent to 8~tons/m$^2$).

We measured the deformation by applying load 
with a press machine to the bottom OEV modules.  
As shown in Fig.~\ref{fig:loadtest},   
the bottom OEV module was fixed on an aluminum spacer 
that has a slope of the same inclined angle of the module 
to keep the upper surface horizontal.
Each module was pressed uniformly up to 1.2 times of the total weight.
The displacement of the top surface was measured with a dial gage. 
Note that in advance of the measurement, we pressed the modules once %beforehand 
to remove the gap between the module and the aluminum spacer.
%Deformations due to loaded weight were checked.

\begin{figure}
\begin{center}
\includegraphics[width=8.7cm]{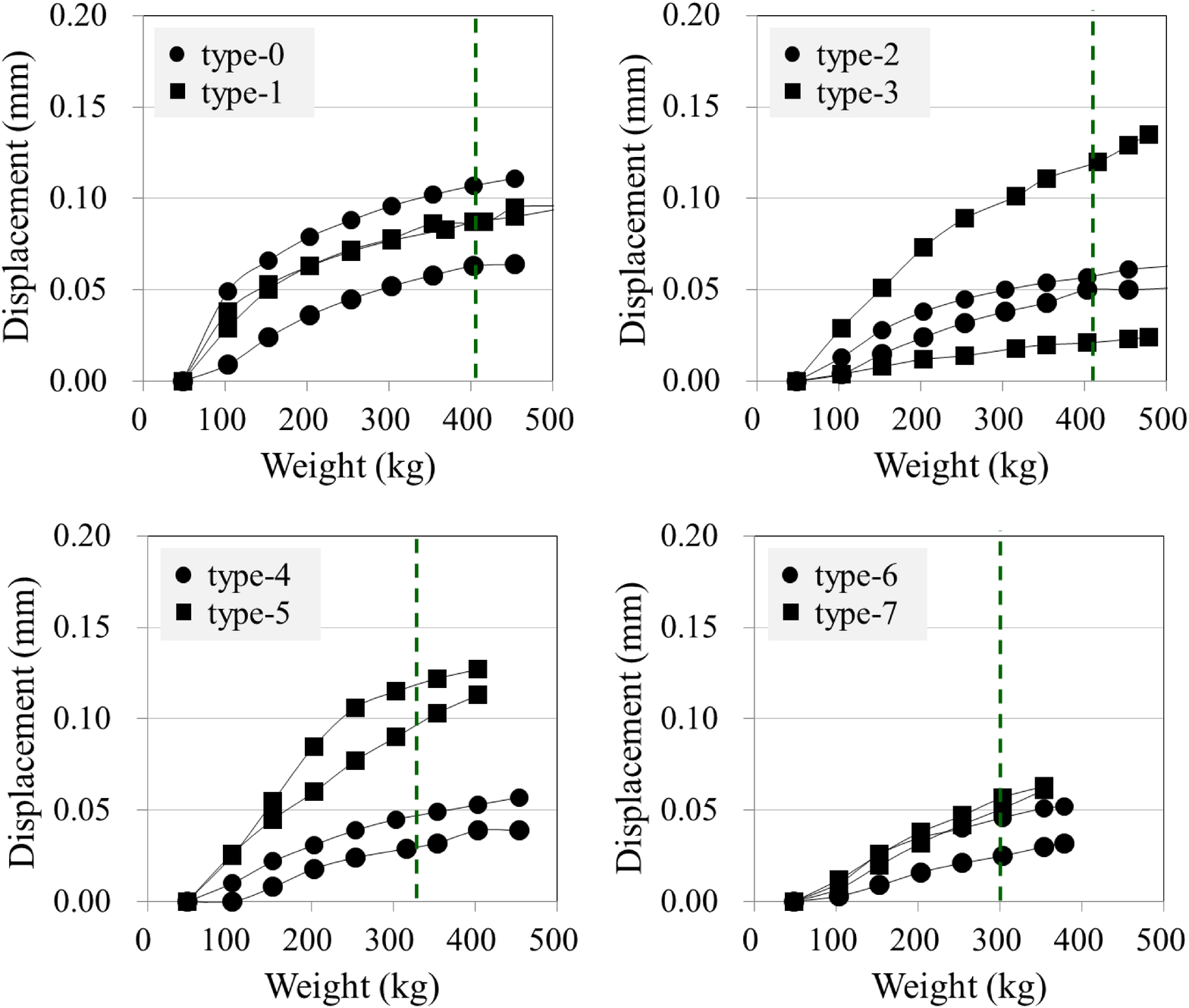}
\caption{Displacement vs.\ applied load of the sixteen bottom OEV modules. 
A vertical dashed line shows the total weight of the CsI crystals stacked on the modules.}
\label{fig:compression}
\end{center}
\end{figure}

Figure~\ref{fig:compression} shows the results of the load test for the sixteen OEV modules 
which will have over 200-kg weight of CsI crystals stacked above them.
Displacements due to the total weight were in the range of  20--120~$\mu$m, and were reproducible\footnote{
Reproducibility was checked with two OEV modules by repeating the load test three times.}.  
The mean and the standard deviation of the displacements   
were 61~$\mu$m and  34~$\mu$m, respectively.
%These results show that the stacking accuracy of the neighboring crystals 
%is less than 200~$\mu$m during the CsI stacking.
These results show that the mechanical robustness of the OEV modules meets the requirement.

%%%%%%%%%%%%%%%%%%%%%%%%%%%%%%%%%%%%
%%%%%%%%%%%%      PMT      %%%%%%%%%%%%%
%%%%%%%%%%%%%%%%%%%%%%%%%%%%%%%%%%%%

\subsection{PMT performance}

The $K^0_L$ decay region of the KOTO detector is 
in vacuum at $10^{-5}$~Pa to eliminate interactions between beam particles and residual gases.  
However, the pressure in the endcap region where PMTs and electronics are located is kept at $\sim0.1$~Pa. 
It had to be verified that no discharge occurred on the electrodes of the PMTs.
Hence, we tested the discharge characteristic of the PMTs used for the OEV counters in vacuum conditions, 
as well as the general characteristics including the gain-voltage relation. 

\subsubsection{Gain} 
\label{subsec:PMTgain}

We evaluated (1) the voltage dependence of the PMT gain and 
(2) the gain fluctuation due to the amplification process from the measured ADC distribution 
by illuminating each PMT with a very weak light source.

\begin{figure}
\begin{center}
\includegraphics[width=7.5cm]{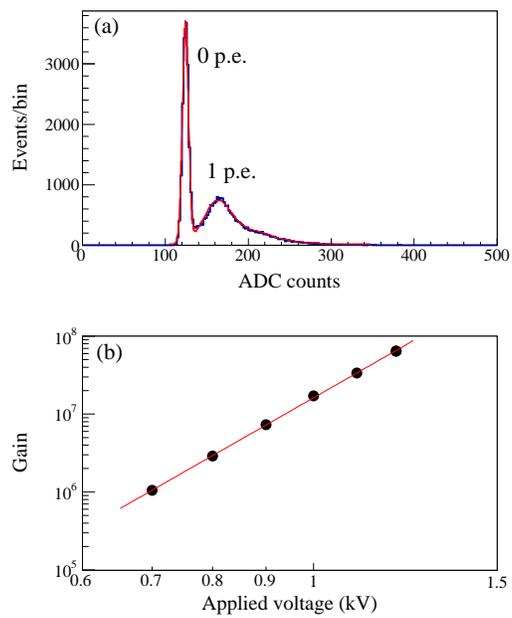}
\caption{(a) ADC distribution of a PMT illuminated with weak light from a green WLS fiber. 
The ADC value corresponding to the single photo-electron peak and its width 
can be estimated by fitting a convolution function of Poisson and Gaussian 
distributions shown as the red curve.  
(b) Voltage dependence of the gain for a typical PMT (Hamamatsu R1924A). 
The straight line shows a fit with the relation $G=KV^\alpha$ (see text). (Color on-line)} %PMT-22
\label{fig:PMTgain}
\end{center}
\end{figure}

In the measurement, a 10-cm-long green WLS fiber (Kuraray Y-11) was placed in front of a PMT.  
The WLS fiber was illuminated with a blue LED (Nichia NSB320BS) whose 
light intensity was adjusted so that the photo-electron (p.e.) yield was 1--3~p.e.\ for each LED pulse 
(see Fig.~\ref{fig:PMTgain}a). 
The signal from the PMT was amplified by 63 times with an amplifier 
(Hamamatsu Photonics C5594~\cite{Hamamatsu}) 
and the output charge was digitized with a 12-bit charge-sensitive ADC (REPIC RPC-022).  
The gate width and the resolution of the ADC were 80~ns and 0.277~pC/ADC count, respectively.

The gain of each PMT ($G$) can be estimated 
from the output charge produced by a single photo-electron ($Q$) as 
$G=Q/e$, where $e$ is the electron charge. 
Figure~\ref{fig:PMTgain}b shows the relation 
between the PMT gain $G$ and the applied voltage $V$~[kV] from 0.7 to 1.2~kV 
with an interval 0.1~kV  for a typical PMT. 
The line shows a fit of $G=KV^{\alpha}$ to the data points, 
where $K$ and $\alpha$ are arbitrary constants and $V$ is in kV.
The means with the standard deviations of $K$ and $\alpha$ 
were $(1.49\pm0.15)\times10^7$ and  $7.71\pm0.09$, respectively. 
The relative spread of the gain among the 44 PMTs ($\sigma_{\rm G}/G$) was 10\% at 0.8~kV.
These parameters can be used to tune the PMT gain. 

The gain fluctuation due to the amplification process 
can be obtained from the width of the single photo-electron peak 
observed in the ADC distribution (see Fig.~\ref{fig:PMTgain}a). 
The width of the single photo-electron peak was $0.48\pm$0.05~p.e., 
averaged over all the PMTs.
This value was utilized to calculate the energy smearing in the Monte Carlo simulation 
as discussed in Section~\ref{subsec:MC}.

\subsubsection{Operation in vacuum}

\begin{figure}
\begin{center}
\includegraphics[width=7cm]{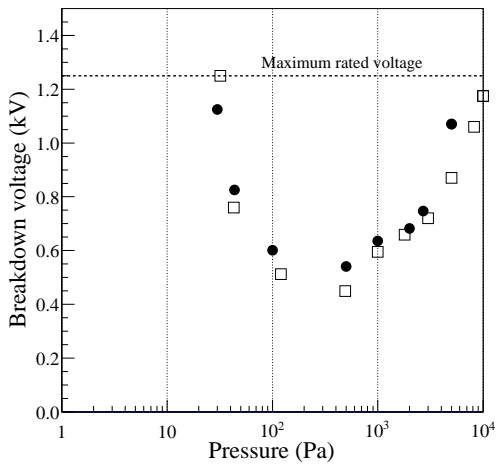}
\caption{
Breakdown voltage as a function of the pressure 
for two randomly selected PMTs (circles and squares). 
There was no indication of discharge in the range of 0.5--10~Pa. 
The dashed line denotes the maximum rated voltage of 1250~V.}
\label{fig:vacuum}
\end{center}
\end{figure}

The manufacturer tested five randomly selected PMTs in vacuum at $10^{-3}$~Pa 
and found no electric discharge on the CW bases for 48~hours. 
%All PMTs were observed with stable output signals. 
The output signal was stable for all PMTs. 
In addition, %assuming 
in order to check the effects of possible 
worse vacuum conditions at the detector commissioning stage,  
we conducted a test at pressures above 0.5~Pa for two randomly selected PMTs. 
As shown in Fig.~\ref{fig:vacuum}, we observed continuous discharge (breakdown) 
in the range of 30 to 10$^4$~Pa  
below the maximum rated voltage of 1250~V. 
Although we occasionally found unstable current in the range 10--30~Pa, 
there was no indication of discharge below 10~Pa. 
These results show that the PMTs operate safely at the normal operating voltages 
for the data taking (750--800~V) under vacuum level of 0.1~Pa,  
which is the typical condition around the endcap region.

%%%%%%%%%%%%%%%%%%%%%%%%%%%%%%%%%%%%
%%%%%%%%%               LIGHT YIELD            %%%%%%
%%%%%%%%%%%%%%%%%%%%%%%%%%%%%%%%%%%%

\subsection{Light yield}

To detect photons having energies above 100~MeV with the required efficiency ($1-\varepsilon<10^{-4}$)~\cite{Ineff2}, 
the minimum detectable energy is 1~MeV in the visible energy\footnote{This low energy threshold is required 
because of  the sampling fluctuations and photonuclear interactions according to the experimental results 
reported in Refs.~\cite{Ineff1,Ineff2}.}. 
If a definite timing information is needed, 
the light yield is required to be, roughly speaking, more than 10~p.e./MeV. 
In this section we discuss the light yield and its position dependence 
along the longitudinal direction obtained with cosmic rays.  

\begin{figure*}
\begin{center}
\includegraphics[width=14cm]{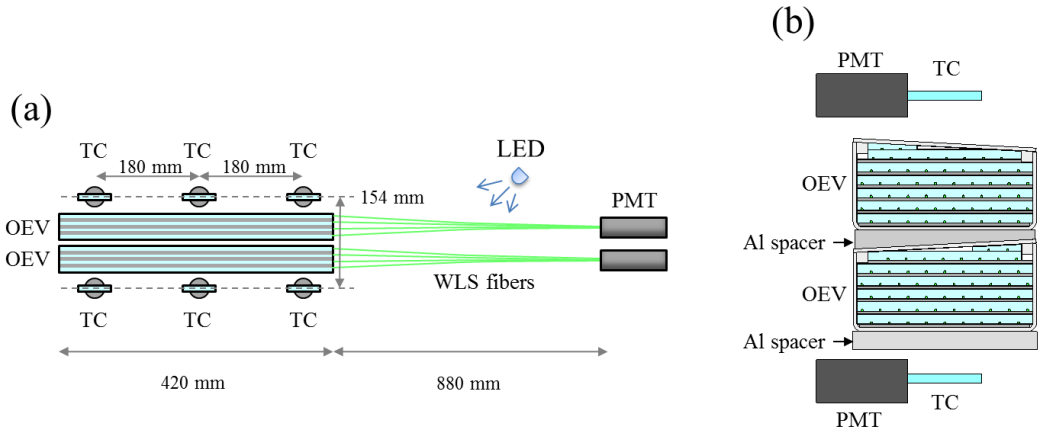}
\caption{
(a) Side view of the setup for the measurement of the light yield and the time resolution with cosmic rays. 
Two OEV modules (type-0) were aligned horizontally and were tested at the same time. 
Three pairs of trigger counters (TC), consisting of $40\times40\times5$~mm$^3$ plastic scintillators, 
were used to detect the passage of cosmic rays. The blue LED that illuminated the WLS fibers 
was employed to measure the single photo-electron peak in the ADC distribution. 
(b) Expanded front view of the setup.  
}
\label{fig:cosmic}
\end{center}
\end{figure*}

\subsubsection{Setup for the light yield measurement}
\label{subsubsec:LYsetup}

The schematic setup for the light yield measurement is shown in Fig.~\ref{fig:cosmic}. 
Two OEV modules of the same type with mirror symmetry    
were stacked horizontally using aluminum spacers.
Three pairs of counters, consisting of 40~mm(length) $\times$ 40~mm(width) $\times$ 5~mm(thickness) 
plastic scintillators, provide the trigger signal for cosmic ray muons passing through the tested OEV modules. 
The cosmic-ray trigger was provided 
by the sum of coincidence signals of the three pairs of trigger counters.  
The trigger counters were placed at equal intervals of 180~mm 
in the longitudinal direction to test the position dependence of the light yield. 
We achieved the light yield measurement with an accuracy of 5\% in half a day.  
The WLS-fiber bundle which was connected to the PMT was illuminated with a blue LED 
(Nichia NSB320BS) as a gain monitor. 

The signals from the OEV modules were amplified threefold with a Phillips Scientific 777~\cite{Phillips} amplifier.
The same voltage of 900~V (with a gain of $(5.2$--$8.0)\times 10^6$) was applied to all PMTs.
The output signal from the amplifier was integrated   
with a 12-bit charge-sensitive ADC, REPIC RPV-171, with a gate width of 120~ns. 
The gains of the PMTs for 1~p.e. were monitored 
at a frequency of 0.5~Hz with weak light from the blue LED.

\subsubsection{Light yield and its position dependence}
\label{subsec:ligghtyeild}

\begin{figure}
\begin{center}
\includegraphics[width=8.5cm]{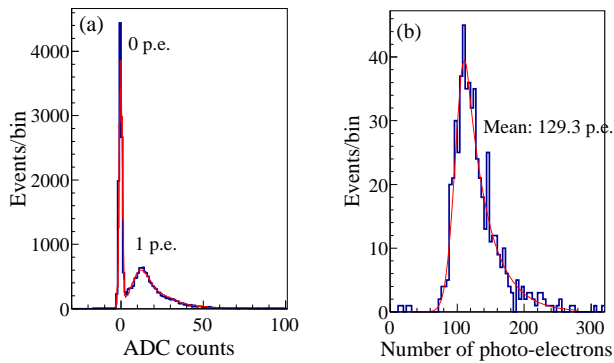}
\caption{(a) ADC distribution of an OEV module (type-0) collected in coincidence with the LED lighting. 
The pronounced peaks around 0 and 12 ADC counts  
correspond to the pedestal and single photo-electron peak, respectively. 
The red curve shows the convolution function of Poisson and Gaussian distributions 
fitted to the ADC distribution.  
(b) The same distribution as (a) collected with the cosmic-ray trigger for 36.7~hours. 
The ADC values have been converted to the number of photo-electrons.
A Gaussian function with an exponential slope (Eq.~\ref{eq:gaustail}) was used to fit the data and 
%mean-value estimation of 
obtain the mean value of 
the number of photo-electrons. (Color on-line)
}
\label{fig:ADCdist}
\end{center}
\end{figure}

Figure~\ref{fig:ADCdist} shows the ADC distributions of an OEV module (OEV-32)   
collected in coincidence with the LED lighting (Fig.~\ref{fig:ADCdist}a) and 
with the middle trigger counters (Fig.~\ref{fig:ADCdist}b) 
for a data collection time of 36.7~hours. 
The position of the single photo-electron peak in Fig.~\ref{fig:ADCdist}a 
was estimated to be $12.1\pm0.1$~counts 
with a convoluted fitting function of Poisson and Gaussian distributions.
The ADC distribution in Fig.~\ref{fig:ADCdist}b shows a minimum-ionization peak 
caused by cosmic ray muons passing through the module.  
The horizontal scale has been converted 
to the number of photo-electrons ($n_{\rm p.e.}$) using the factor of 12.1~counts/p.e. obtained earlier.

To obtain the mean value of the light yield, 
we fit the data with the function:    
\begin{linenomath}
\begin{equation}
  f(x)=  \begin{cases}
                  A \exp \left[\displaystyle{ \frac {-(x-x_{\rm p})^2} { 2 \sigma^2 }}\right] &  ( x \le x_{\rm b})  \\
                  B \exp [ -\lambda (x-x_{\rm b}) ] & ( x > x_{\rm b})
                   \end{cases}           
, \label{eq:gaustail} 
\end{equation}
\end{linenomath}
where $A$ denotes the peak value of the Gaussian at $x=x_{\rm p}$ and $B$ is  
the value at the boundary of the two functions ($x_{\rm b}$). 
Note that the parameters $B$ and $\lambda$ are determined 
under the continuity condition of the two functions at the boundary $x_{\rm b}$.  
There are four free parameters: 
$A$, $x_{\rm p}$, $x_{\rm b}$, and $\sigma$ in the fitting calculation. 
For the OEV module used in Fig.~\ref{fig:ADCdist}b,  
the mean value was calculated to be ($129.3\pm1.5$)~p.e. from the first moment of the fitted function. 
% PDG                   de/dx         density         de/dl
%                      MeV/(g/cm2)      g/cm3       MeV/cm
%        Polystyrene  1.936            1.06            2.052
%        acrylic         1.929　　　　　  1.19            2.296
%        MS              1.935             1.073          2.078
On the other hand, we estimated the average energy deposit in the plastic scintillators 
to be 6.24~MeV, assuming that the energy deposit and the average path length of cosmic rays  
in the scintillator are 2.08~MeV/cm and 30.0~mm, 
respectively\footnote{The energy deposit in the plastic scintillator based on the MS resin, 2.08~MeV/cm,   
was calculated based on the weighted average of $dE/dx|_{\rm min}$ for polystyrene and acrylic, as described in Ref.~\cite{Klp0p0}. 
The density of the MS resin was assumed to be 1.075~g/cm$^3$.}.
% by YOKOTA       
% bottom type 0  30.3 mm   assuming plastic scintillator thickness of 5mm
%                 angle correction  1.02
%                       30.3*4.85/5*1.02=30.0 mm
%  
Thus, the mean light yield of this module was ($20.7\pm0.2$)~p.e./MeV.

\begin{figure}
\begin{center}
\includegraphics[width=7.5cm]{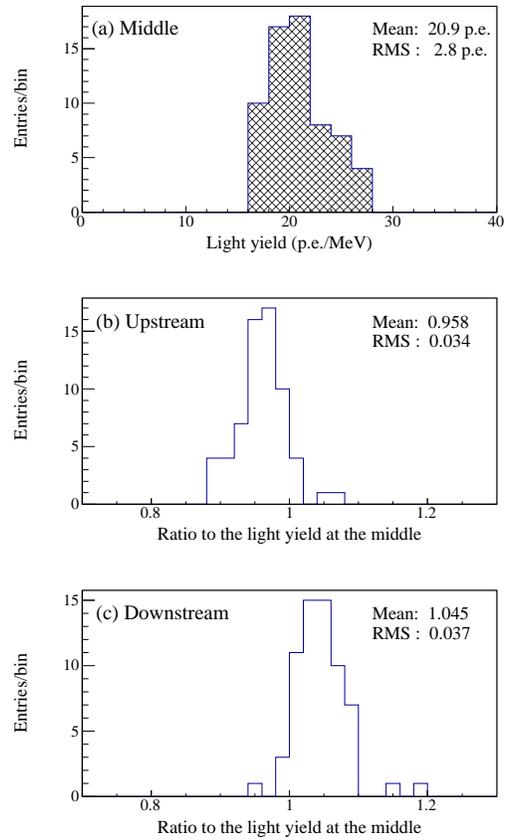}
\caption{
(a) Distribution of the mean light yields collected with the cosmic rays passing through 
the middle trigger counters for all the 64 OEV modules. The mean light yield 
taken with (b) the upstream and (c) the downstream trigger counters 
were plotted as the ratio to the middle one. 
The mean and the root mean square (RMS) of the distributions are also reported.    
}
\label{fig:LY}
\end{center}
\end{figure}

Figure~\ref{fig:LY}a shows the distribution of the mean light yields for all the modules  
collected with the middle trigger counters.
The mean and the standard deviation of 
the light yield for 64 modules were 
20.9~p.e./MeV and 2.8~p.e./MeV, respectively. 
The minimum light yield among the 64 modules was 
16.2~p.e./MeV for a type-2 module   
and is sufficient to detect an energy deposit of 1 MeV. 
The result was consistent with a previous study in which the light yield 
was estimated to be 17~p.e./MeV~\cite{MB} with similar detector components (MS-resin extruded scintillator,  
Y-11 fibers and Hamamatsu R329, and a standard bi-alkali PMT).

To check the position dependence along the fiber direction, we plotted 
the ratios of the light yield at the upstream and the downstream ends to the yield measured at the middle, 
as shown in Fig.~\ref{fig:LY}b and \ref{fig:LY}c. 
The dependence shows a similar tendency among all the OEV modules: 
the yield is lower near the front face and greater near the rear side.  
%The differences of the ratios were 5\%, 
The light-yield differences at the both ends were 5\%, 
corresponding to the effective attenuation length of 350~cm 
mainly due to the light attenuation in the WLS fibers.   
The minimum light yield of the three measured points 
among all the OEV modules was 15.4~p.e./MeV at the upstream for a type-2 module. 
They are well within the requirement.

%%%%%%%%%%%%%%%%%%%%%%%%%%%%%%%%%%%%
%%%%%%%%%               TIME RESOLUTION                      %%%%%%
%%%%%%%%%%%%%%%%%%%%%%%%%%%%%%%%%%%%
   
\subsection{Time resolution}

In the KOTO experiment, the time resolution of veto counters 
determines the time window for the photon veto,   
which is directly related to the acceptance loss of the \klpnn\ signal due to accidental hits.
The required time resolution for the OEV counters is less than 10~ns\footnote{This requirement 
is needed in order to keep the acceptance loss below 1\% under an accidental rate of 100~kHz for the sum of all the OEV counters. 
A veto time window of $\pm 5\sigma$ ($=100$~ns) is assumed.}. 
In this section, we evaluate the time resolution of the OEV counters and its position dependence 
with two different conditions: 
(1) in a bench test using VME TDCs with the same setup as the light yield measurement and 
(2) in the real condition using 125-MHz waveform digitizers~\cite{FADC} used in the readout of the KOTO experimental setup.

\subsubsection{Time resolution evaluated in the bench test}

The detector setup in the bench test was the same as the one used for 
the light yield measurement described in Section~\ref{subsubsec:LYsetup}.
One of the output signals 
was processed with a constant-fraction discriminator (CFD) Phillips Scientific 715. 
We set the CFD threshold at $-150$~mV which corresponds to $17.3\pm2.6$~p.e.
%depending on the PMT gain.   
The time difference between the trigger signal and the output signal from the CFD 
was digitized with a 12-bit CAEN V775 TDC~\cite{CAEN} with a full range of 150~ns.   
The timing of the trigger signal for each pair of trigger counters
was determined by the bottom counters.

In the data analysis, we obtained the time resolution of each OEV module
by fitting a Gaussian to the time-difference distributions
between the bottom trigger counters and each OEV module.
The timing spreads due to the three sets of trigger counters 
were measured to be 0.162~ns (upstream), 0.206~ns (middle), and 0.218~ns (downstream)\footnote{
These values were obtained from widths of the time-difference distributions 
of each pair of trigger counters assuming an equal time resolution for both the trigger counters.}.  
They were subtracted quadratically.

\begin{figure}
\begin{center}
\includegraphics[width=7.5cm]{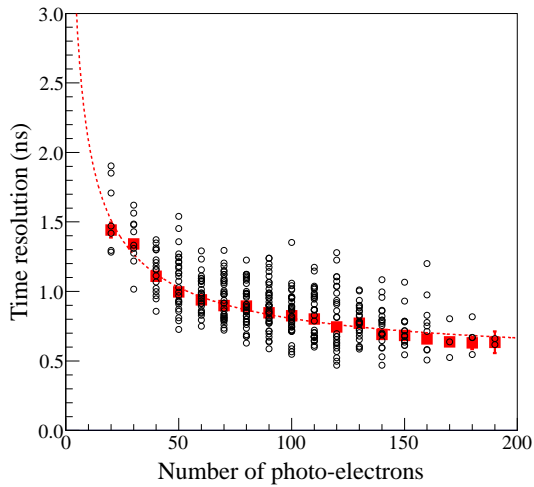}
\caption{
Time resolution of the 64 OEV modules measured with the middle trigger counters 
as a function of the number of photo-electrons (open circles).  
Only the data having relative uncertainties lower than 20\% are plotted.  
The weighted-mean values of the time resolutions for different $n_{\rm p.e.}$ bands
are also shown (filled red squares).     
The dotted curve indicates the fit function of Eq.~\ref{eq:treso} in the text, 
where $a=6.4\pm0.2$~ns and $b=0.49\pm0.02$~ns. (Color on-line)}
\label{fig:time_reso}
\end{center}
\end{figure}

The scatter plot in Fig.~\ref{fig:time_reso} shows the time resolutions 
in the middle of the 64 OEV modules for different light yields.
Note that the resolutions were calculated for each module with an interval of 10 photo-electrons   
because the shape and the range of the light-yield distributions (see Fig.12b for example) 
varied depending on the module types\footnote{The data points at lower light yields are dominated 
by the modules having thinner scintillator layers due to the fact that 
the energy deposit in the scintillators depends on the path length of cosmic rays.  
The average scintillator thickness ranges from 7.7~mm (for type-2) to 33.4~mm (for type-8). }. 
In total, 365 data points were plotted after removing 
the data with high relative uncertainty ($>20\%$). 
%The data points at lower light yields are dominated 
%by the modules having thinner scintillator layers due to the fact that 
%the energy deposit in the scintillators depends on the path length of cosmic rays.  
%The average thickness of the scintillators ranges from 7.7~mm (for type-2) to 33.4~mm (for type-8).  
In order to check that the resolution is dominated by the statistical fluctuations in the number of photo-electrons, 
the weighted mean of the resolution 
among the data with the same light-yield bin was calculated, 
showing a monotonic decrease as a function of the light yield (filled squares in Fig.~\ref{fig:time_reso}).
%##############################################
% YOKOTA   7.8 mm - 33.0 mm
%                 7.8*(4.85/5)*1.02=7.7 mm
%                 33.8*(4.85/5)*1.02=33.4 mm
%##############################################
Assuming Poisson statistics for the number of photo-electrons, 
we fitted the following empirical function,                                                                                    
\begin{linenomath}
\begin{equation}
\sigma_{\rm t} =\frac{a} {\sqrt{n_{\rm p.e.}}} \oplus b , \label{eq:treso}
\end{equation}
\end{linenomath}
to the weighted means, where $a$ and $b$ are free parameters 
and $\oplus$ denotes the square root of the quadratic sum. 
%The fit results give  
The parameters determined from the fit are 
$a = 6.4 \pm 0.2$~ns and $b = 0.49 \pm 0.02$~ns.
This relation determines the typical value of the time resolution 
to be 1.5~ns for energy deposit of 1~MeV, which is obtained by using the mean light yield  
20.9~p.e./MeV described in Section~\ref{subsec:ligghtyeild}.
For individual modules, the time resolutions at 1~MeV may distribute around the typical value. 
Nevertheless, the requirement of less than 10~ns is expected to be met from the overall trend.

\begin{figure}
\begin{center}
\includegraphics[width=7.5cm]{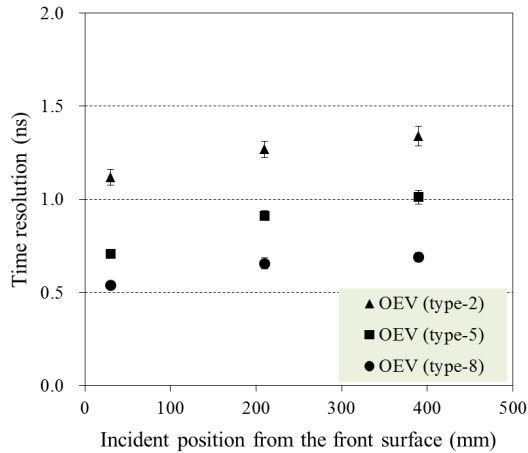}
\caption{
Position dependence of the time resolution for three bottom OEV modules. 
%The path length of cosmic rays in the scintillators of each type of modules
%is different. Therefore the time resolution in this plot is also different.
The mean energy deposits in the scintillators are estimated to be 
1.7~MeV, 3.6~MeV, and 6.3~MeV 
for type-2, type-5, and type-8, respectively. 
}
\label{fig:time_reso_type}
\end{center}
\end{figure}

The typical longitudinal-position dependence of the time resolution 
is shown in Fig.~\ref{fig:time_reso_type} for three types of modules (type-2, 5, and 8).   
The path length of cosmic rays in the scintillators of each type of modules
is different. Therefore the time resolution in this plot is also different. 
Although the light yield was 5\% larger at the downstream position as discussed in the previous section,  
the time resolution was 13\% worse at this location.  
This tendency originates from the time spread of re-emitted photons from the WLS fibers.  
In reality, the optical-path differences in WLS fibers 
between reflected photons at the front surface of each module 
and the direct photons toward the PMT 
are 6~cm, 42~cm, and 78~cm at the three positions of the trigger counter. 
With the propagation velocity of 17.5~cm/ns~\cite{MB}, 
these cause the time difference of 0.3~ns, 2.4~ns, and 4.5~ns between the direct and reflected photons.
As a result, the photon-detection-time spread is a combination of  
the optical-path difference and the decay time of the WLS fiber ($\tau=6.8$~ns~\cite{MB}).
Although the position dependence of the time resolution is not significantly large, 
it should be taken into account for the setting of the veto time window.

\subsubsection{Time resolution evaluated in the FADC readout}

In the KOTO experiment, timing information of the OEV counters 
is extracted with the analysis of the waveforms recorded by the 125-MHz flash ADC (FADC)~\cite{FADC}.
Here, we will discuss the time resolution measured with the FADC readout.

\begin{figure}
\begin{center}
\vspace{8cm}
\includegraphics[width=6.5cm]{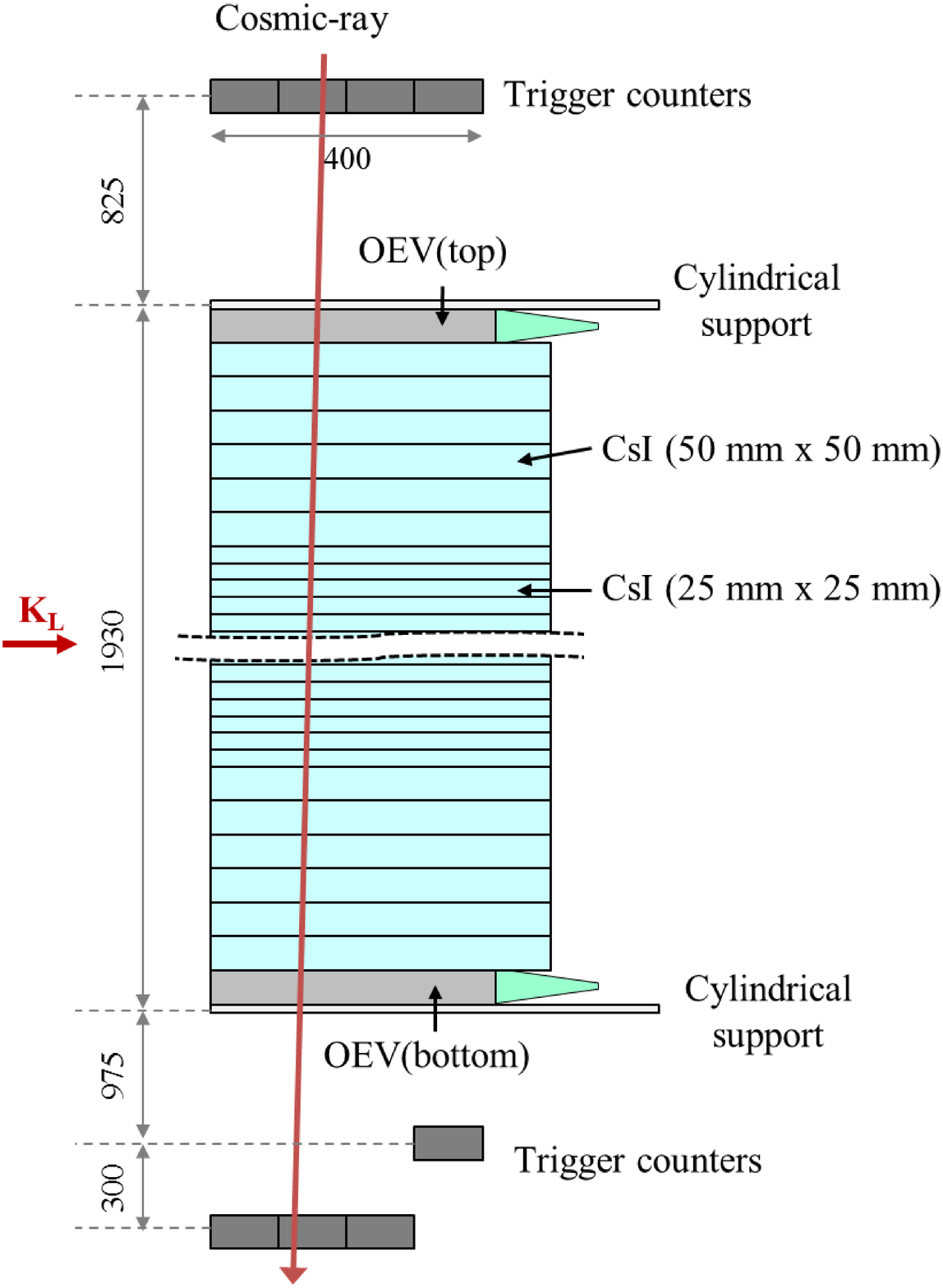}
\caption{Setup for the time resolution measurement of 
OEV-10 and OEV-33, located at the top and the bottom of the endcap, 
with cosmic rays. 
Four pairs of trigger counters are set horizontally at both sides of the endcap 
with orthogonal direction to the OEV counters. 
Each trigger counter 
%which is made of plastic scintillator (10~cm wide and 5~cm thick and 200~cm long), 
is read out from both ends with two PMTs. 
With this setup of the trigger counters, the active length of the OEV counters (42~cm) 
is divided into four 10-cm regions.}
\label{fig:cosmic_setup}
\end{center}
\end{figure}

The measurement was performed for two OEV counters, OEV-10(top) and OEV-33(bottom), 
seven months after the installation of the OEV counters in February 2011. 
Four pairs of trigger counters %having the length of 200~cm 
200~cm long were set horizontally 
at the top and bottom of the OEV counters with a vertical distance of about 4~m   
as shown in Fig.~\ref{fig:cosmic_setup}. 
%The size of 
Each trigger counter was 10~cm wide and 5~cm thick. 
By requiring a coincidence in a pair of 
%the top and bottom 
trigger counters, one can select 
%signals of the cosmic ray passing vertically through four 10~cm-regions 
cosmic rays passing vertically through the corresponding four regions  
of the OEV counters.  
The average energy deposits 
are estimated to be 6.5~MeV and 6.2~MeV for OEV-10 and OEV-33, respectively. 
We applied the same voltage of 850~V to the PMTs. 

\begin{figure}
\begin{center}
\includegraphics[width=7.5cm]{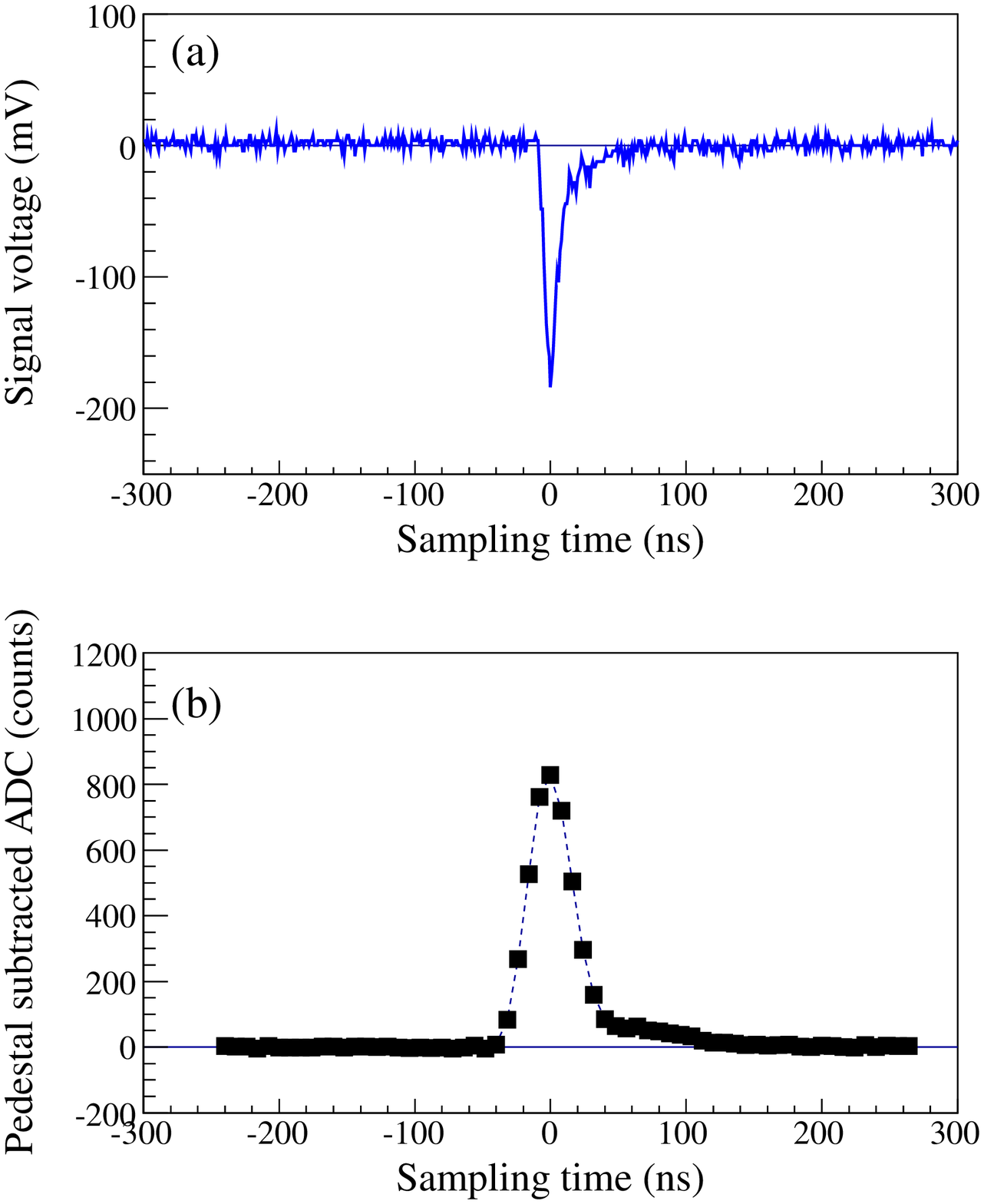}
\caption{Typical waveform of a cosmic-ray signal recorded  
(a) with a 1-GHz digital oscilloscope (Tektronix DPO4104~\cite{Tektro}) and  
(b) with the 125-MHz FADC after the 10-pole low-pass filter.
The closed squares show 64 sampling points of the FADC while  
the dashed lines connect two adjacent sampling points.}
\label{fig:pulse_shape}
\end{center}
\end{figure}

%In common with the KOTO data acquisition (DAQ) system, 
Analog signals from the OEV counters were digitized with the 125-MHz FADC.  
A 10-pole low-pass filter with the function of stretching the pulse shapes to a Gaussian form 
was utilized to accommodate the relatively long sampling interval (8~ns) 
comparing to the pulse width of the OEV counters. 
The filter also suppresses high-frequency noise. 
Figure~\ref{fig:pulse_shape} shows the typical pulse shapes of the raw and the filtered signals. 
Timing was defined as the rising time at half maximum from the fitted Gaussian function.

\begin{figure}
\begin{center}
\includegraphics[width=7.5cm]{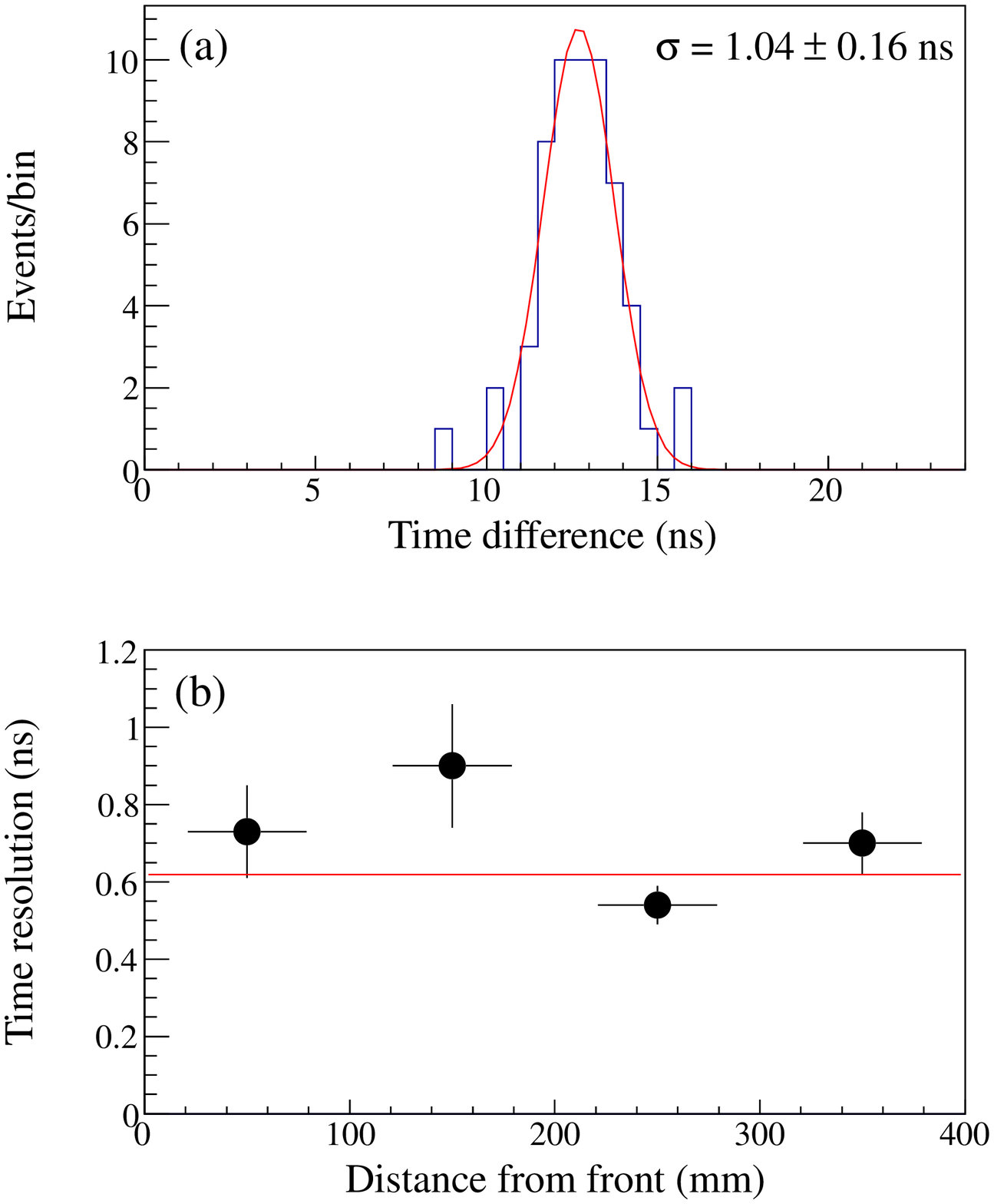}
\caption{(a) Time-difference distribution between OEV-10 and OEV-33 at the most upstream region. 
The curve shows the fitted Gaussian. 
(b) Position dependence of time resolution for OEV-10 and OEV-33 
obtained from the time difference distribution. The line shows the weighted mean of four regions.}
\label{fig:timereso_posi}
\end{center}
\end{figure}

The time resolution was estimated from the width ($\sigma$) of the time-difference distribution 
between the top and bottom OEV counters (see Fig.~\ref{fig:timereso_posi}a). 
By assuming equal time resolution for both OEV counters in each region, 
the resolution of each module is thus $\sigma/\sqrt{2}$. 
Figure~\ref{fig:timereso_posi}b shows the time resolution for each region.%the four divided regions.    
Since there is no significant position dependence within this rather large error, 
we concluded that the time resolution measured with the FADC 
was $0.62\pm0.04$~ns from the weighted mean of the four data points. 
This result is consistent with the mean time resolutions evaluated at the three positions  
measured with the bench test described in the previous section.  
%%%%%%%%%%%  weighet means of the three positions
This implies that the readout method with the low-pass filter and the 125-MHz FADC 
does not worsen the timing performance significantly.

\section{Energy calibration with cosmic ray muons}
\label{sec:5}
%%%%%%%%%%%%%%%%%%%%%%%%%%%%%%%%%%%%
%%%%%%%%          CALIBRATION       %%%%%%%
%%%%%%%%%%%%%%%%%%%%%%%%%%%%%%%%%%%%

An energy calibration method with cosmic ray muons 
was developed for the data-taking period. 
Because of the trapezoidal shapes and the two different directions of stacking layers of the OEV counters 
(horizontal and vertical), the path length of the muon in the scintillators of each counter 
depends upon its incident position and angle. 
The energy distribution in some OEV counters shows no peak structure for cosmic ray muons.  
A simple calibration method based on the minimum-ionization peak cannot be applied. 
A technique was exploited to obtain 
the calibration constants (ADC counts/MeV) for all the counters 
by fitting the ADC distribution generated with a Monte Carlo (MC) simulation 
to the experimental data.

\subsection{Experimental conditions}
\label{subsec:cosmi_setup}

In the data-taking period, 
calibration data were recorded for 2~seconds during 
the beam-off period out of the 6-second acceleration cycle\footnote{
The acceleration cycle in the slow-extraction mode at J-PARC   
consists of a beam-on period of 2 seconds and a beam-off period of 4 seconds. }.
We analyzed calibration data taken in about three days in May of 2013  
with a beam power of 24~kW (3$\times 10^{13}$ protons/pulse).
%%%%%%### 23.8e3/(30e9*1.6e-19/6.)
The PMT gains of the OEV counters were tuned to $2\times10^6$. 
All the PMTs operated stably
without any discharge during the data taking with vacuum less than 0.1~Pa. 
Because of the limited number of sensors, temperature was monitored for only 
two PMTs (OEV-10 and OEV-22) with thermocouples. 
The means of the recorded temperatures were 
$26.5^\circ$C and $32.0^\circ$C for OEV-10 and OEV-22, respectively. 
Their variations were within $\pm0.5^\circ$C \footnote{
The temperature for OEV-10 was lower because this counter was 
just behind a thick stainless steel plate, where the heat radiated 
from many CsI calorimeter PMTs was shielded.}.

The data acquisition system for the calibration data was triggered 
when the total energy deposit in the CsI calorimeter was greater than 600~MeV. 
This rather high energy threshold requires cosmic ray muons to have long track length 
in the calorimeter ($>77$~cm) and reduces backgrounds due to low-energy showers.  
As estimated with 125 event samples by a careful eye scan of the event displays, 
more than 98\% of the triggered events were originated from cosmic-ray-muon tracks.   
The trigger rate was 40~Hz   
%%% from RUN 16793
%%% Trig-6 CsI(TotalEt) offspill: prescale factor 1, Counts/Spill=159.68 
%%% XXXX 159.68/4sec = 40 Hz <- This is incorrect.
%%% 159.68/2sec = 80 Hz %% MOD by toru 2013.12.2 <- this is also incorrect(laser data included)
%%% from RUN 16680
%%% 94149 events were triggered -> 47698 events remain after the laser data rejection
%%% 47698/(601 spill * 2 sec) = 39.7 Hz 
and $4.8\times10^4$ triggered events ($=80~{\rm triggers/cycle}\times600~{\rm cycles/h}$) 
were accumulated in one hour of data taking.
The 125-MHz FADCs were used for signal digitization.

In the data analysis, we obtained the pulse area by integrating the pedestal-subtracted 
FADC values at all the sampling points\footnote{The number of sampling points was 64 as shown in Fig.~\ref{fig:pulse_shape}b.
The pedestal was set by the mean value of the first 10 sampling points.}.  
\begin{figure}
\begin{center}
%%%%%%% 2013 May Physics 10run  GT 1500ADC counts
\includegraphics[width=8.7cm]{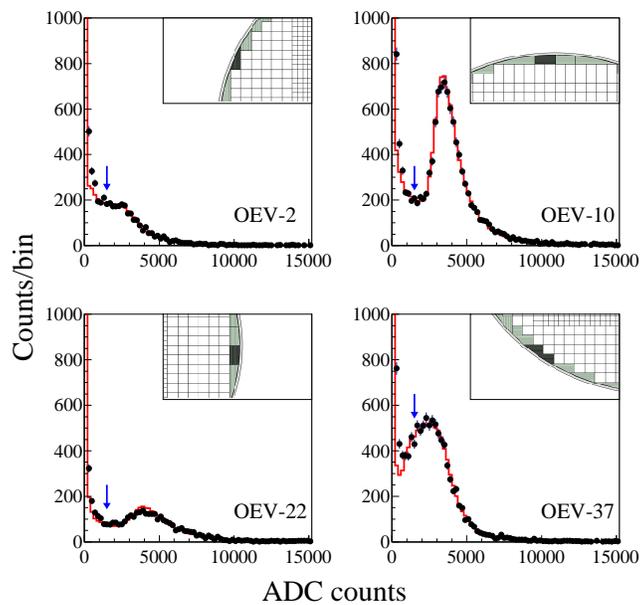}
\caption{Integrated ADC distributions for four typical OEV counters 
after 10 hours of data taking (black points).  
The location of each counter is also illustrated in each plot.  
The histograms show the ADC distributions generated with the MC simulation 
which is described at the end of section~\ref{sec:calib_meth}.
The MC distributions are scaled to the data in the range of ${\rm ADC}>1500$ counts,  
which is indicated by the blue arrows. (Color on-line)}
\label{fig:bestfit}
\end{center}
\end{figure}
The black points in Fig.~\ref{fig:bestfit} show the integrated ADC distributions 
for four typical OEV counters after 10 hours of calibration-data taking.  
The red histograms represent MC distributions to be detailed later.
The distribution for OEV-10, which is located at the top part of the endcap, 
shows a pronounced peak at around 3500 ADC counts   
because the path length of the cosmic ray muons 
in the scintillator plates is nearly constant. 
On the other hand, OEV-2, which is located at the side region, 
has just a small bump-like structure 
since the path length has a large variation 
due to the triangular shape of the module and the vertical stacking direction.

\subsection{Simulation of cosmic rays}
\label{subsec:MC}

A MC simulation of the KOTO detector based on Geant4~\cite{Geant4} was used to calculate 
the energy deposit in each OEV counter. 
The positional difference from the actual setup in the geometric definition 
is expected to be within 1~mm, originated from the deformation of the cylindrical support structure of the endcap.
Radiation-shielding-concrete blocks (2.3~g/cm$^3$) covering the whole setup were also included.   
The shield is at least 1~m thick with a realistic geometry\footnote{
The material definition of the concrete was based on the NIST database~\cite{NIST}.}.

In the simulation, the momentum ($p_\mu$) and zenith angle ($\theta$) of the cosmic ray muons 
were generated based on the following distribution of the muon intensity 
\begin{eqnarray}
I(p_\mu,\theta) &=& \cos^3\theta\cdot I_{\rm V}(p_\mu \cos\theta ),  
\end{eqnarray}
where $I_{\rm V}$ denotes the vertical muon intensity as a function of vertical component of $p_\mu$~\cite{Reyna}. 
This formula reproduces experimental data observed at ground level in the range of $p_\mu>1$~GeV$/c$. 
Considering the high trigger threshold and energy loss in the shielding concrete ($>390$~MeV), 
we can safely neglect the cosmic ray muons below 1~GeV$/c$.
Muons were generated from the air above the shielding concrete 
so that incident positions at the central height of the detector were distributed uniformly in the horizontal plane 
with an area of 9.0~m (along the beam direction) $\times$ 7.6~m (perpendicular to the beam direction). 
By limiting the source area, simulation time can be reduced,  
%without a significant loss of primarily generated muons, 
in particular for muons with small zenith angles.
Energy deposit in each counter was smeared using random numbers generated 
based on photo-electron statistics\footnote{We generated the number of photo-electrons ($n_{\rm p.e.}$) 
in each counter based on the Poisson distribution. The mean value of the distribution 
($\overline{n_{\rm p.e.}}$) was obtained from the energy deposit in the scintillators ($E_{\rm dep}$) and 
the average light yield of all OEV modules ($Y= 20.9$~p.e./MeV) with $\overline{n_{\rm p.e.}}=Y\cdot E_{\rm dep}$. }, 
the fluctuation of the dynode amplification\footnote{The number of photo-electrons was smeared again 
with a Gaussian of width $\sigma=0.48\sqrt{n_{\rm p.e.}}$ by taking the fluctuation of the dynode amplification 
of the PMTs into account as described in Section~\ref{subsec:PMTgain}.}, 
and electrical noises evaluated with actual measurement.

\subsection{Calibration method}
\label{sec:calib_meth}

\begin{figure}
\begin{center}
\includegraphics[width=7.5cm]{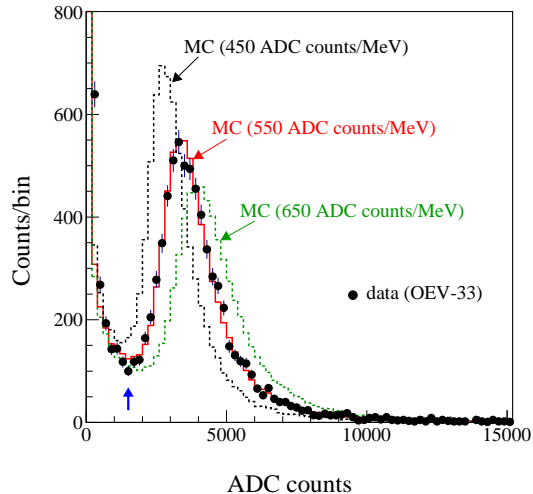}
\caption{
ADC distribution for OEV-33 after 10 hours of calibration-data taking (black points). 
Three histograms show the ADC distributions generated with 
different conversion factors. 
The MC distributions are scaled to the data 
in the range of ${\rm ADC}>1500$ counts,  
which is indicated by the blue arrow. (Color on-line)
}
\label{fig:MC_charge}
\end{center}
\end{figure}

We generated ADC distributions in the MC simulation for each OEV counter, 
assuming various conversion factors between energy deposit and ADC counts 
in the range from 400 to 1000~ADC counts/MeV (Fig.~\ref{fig:MC_charge}).
The generated MC distributions were scaled to the experimental  
distribution in the range above 1500 ADC counts. 
The threshold was set to remove accidental noise at low energy. 
In Fig.~\ref{fig:MC_charge},  
the optimum conversion factor is expected to be 550 ADC~counts/MeV.

In order to find the best value of the conversion factor, 
we introduced the following $\chi^2$ value for the histograms: 
\begin{eqnarray}
\chi^2&=&\sum_{i}^{N_{\rm bin}}\frac{(n_{{\rm data},i}-n_{{\rm MC},i})^2}{\sigma_{{\rm data},i}^2+\sigma_{{\rm MC},i}^2},
\end{eqnarray}
where $n_{{\rm data},i}$ and $\sigma_{{\rm data},i}$ denote 
the counts and the error of the $i$th bin for the data,  
$n_{{\rm MC},i}$ and $\sigma_{{\rm MC},i}$ are those for the MC distribution,  
and $N_{\rm bin}$ represents the total number of bins in the summation\footnote{
In the summation, data points with $n_{{\rm data},i}<4$ were excluded; thus 
the value of $N_{\rm bin}$ depends on the statistics of data. 
Typical $N_{\rm bin}$ values were 66 for OEV-10 and 42 for OEV-2, with 10 hours of data. 
}. 
\begin{figure}
\begin{center}
\includegraphics[width=8.7cm]{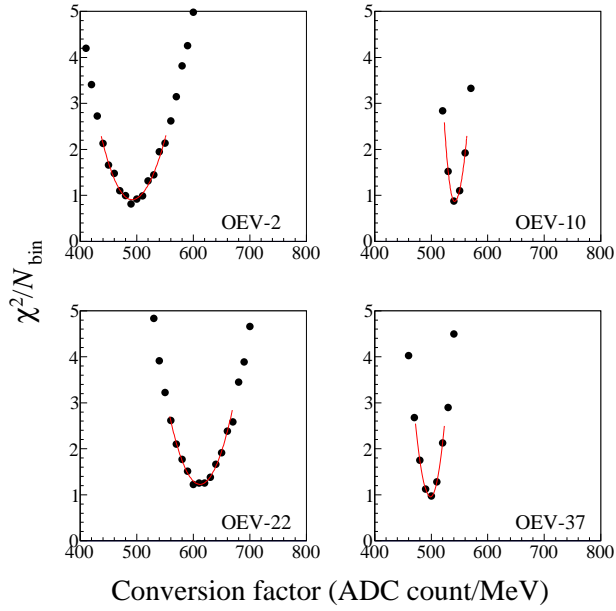}
%%%%%%%%%%2013 May Physics 10run  GT 1500ADC counts
\caption{
$\chi^2/N_{\rm bin}$ as function of the conversion factor 
for the same OEV counters shown in Fig.~\ref{fig:bestfit}. 
$\chi^2/N_{\rm bin}$ reduces when the degree of consistency 
between data and MC distributions becomes high, 
and approaches unity if the distributions are consistent within errors.
The curves show asymmetric parabola functions fitted to each plot around the minimum (see text).}
\label{fig:chisq}
\end{center}
\end{figure}
Figure~\ref{fig:chisq} shows the $\chi^2/N_{\rm bin}$ values 
as a function of the conversion factor for the same OEV counters shown in Fig.~\ref{fig:bestfit}.  
Asymmetric parabola functions fitted to these plots,  
\begin{linenomath}
\[  y=  \begin{cases}
A (x-x_0)^2+y_{\rm min} &  ( x \le x_0)  \\
B (x-x_0)^2+y_{\rm min} & ( x > x_0) \end{cases}     
,\] 
\end{linenomath}
have the minimum values close to unity at 500--700~ADC counts/MeV, 
which indicated that the MC simulation worked properly and well reproduced the experimental data. 
The calibration constants obtained here are in agreement with 
expectations from signal-output charge and the dynamic range (2~V) of the 14-bit FADC.

\subsection{Discussion}

ADC distributions generated with the MC simulation   
are superimposed on the experimental data as shown in Fig.~\ref{fig:bestfit}.
The experimental distributions are in good agreement with the MC distributions. 
The minimum values of  $\chi^2/N_{\rm bin}$ ranged from 0.7 to 2.0 for all the OEV counters. 

The sharpness of the parabola in Fig.~\ref{fig:chisq}
represents the precision of estimates for the calibration constant.  
It depends on both the statistics of the data and the shape of the ADC distribution. 
A quantitative estimate of the calibration error was 
the intersection distance at the minimum $\chi^2$ plus 1 (confidence interval of 68\%). 
\begin{figure}
\begin{center}
\includegraphics[width=8cm]{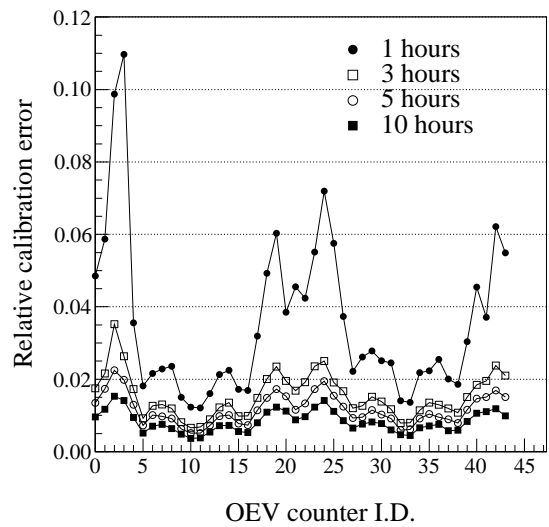}
\caption{Calibration errors for all OEV counters estimated with different 
amounts of data-taking time: one hour (full circles), three hours (open squares),  
five hours (open circles) and 10 hours (full squares). 
The relation between the ID number of the OEV counters and the actual location 
of the counters is shown in Fig.\ref{fig:Endcap}.
}
\label{fig:calib_error}
\end{center}
\end{figure}
Figure~\ref{fig:calib_error} shows the relative calibration errors 
for all the counters with different data-taking time.  
We found the smallest errors for OEV-10, 11, 32, 33 (type-0 modules located at the top and bottom part), 
which showed pronounced peaks in the ADC distributions. 
On the other hand, large calibration errors were obtained 
for OEV-2, 3, 19, 24, 42 (counters located near the side region), 
which showed small bump-like structures in the ADC distributions.
The calibration error after one-hour data taking was estimated to be 1\% 
in the best case and 11\% in the worst case.
Even in the worst case, the error can be easily reduced to be 3\% 
by collecting data for three hours.  
The calibration method gives an accuracy of a few percent for various ADC-distribution shapes.

\begin{figure}
\begin{center}
\includegraphics[width=9.0cm]{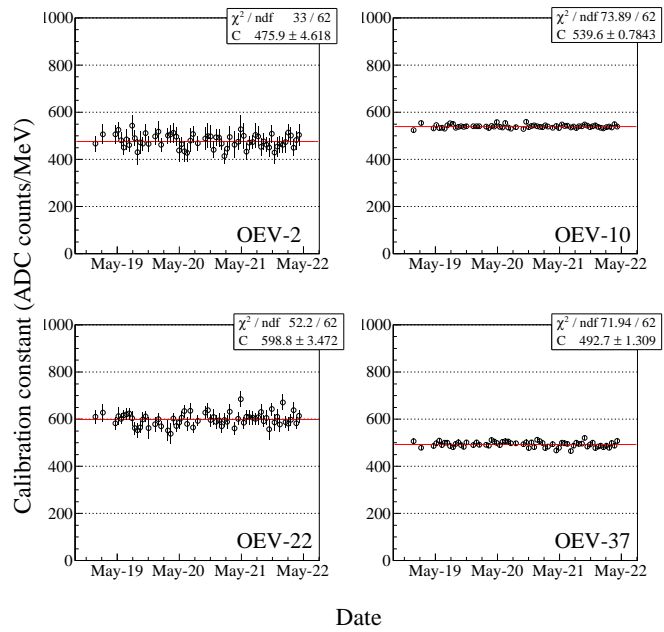}
\caption{
Time variation of the calibration constants evaluated from one-hour data taking 
for the OEV counters shown in Fig.~\ref{fig:bestfit}. 
The larger error bars for OEV-2 and OEV-22 
are attributed to the relatively large calibration error among the OEV counters 
(see Fig.~\ref{fig:calib_error}).
The straight lines show the constant functions fitted to the plots, 
and $C$ is a free parameter of the fit. 
}
\label{fig:gain_variation}
\end{center}
\end{figure}

Stability of the light yield can be monitored by checking 
the time variation of the calibration constant. 
Figure~\ref{fig:gain_variation} shows the evaluated variation from one-hour data taking for four OEV counters.  
Light yields were stable during the data collection period (three days). 
This was also true for the other OEV counters (the range of reduced $\chi^2$ was 0.53--1.59).
Gain variation of the PMTs was thought to be well below the sensitivity limit of the calibration. 
For example, gain variation coming from temperature change 
is expected to be small ($<0.2$\%)\footnote{The temperature coefficient of the gain is typically $-0.31$\%/$^\circ$C 
for Hamamatsu R1924A~\cite{Hamamatsu_p}.} 
since  the temperature variation was within $\pm0.5^\circ$C in this period. 
On the other hand, other factors of light yield variation, such as aging deterioration of light transmission 
in the scintillators and the WLS fibers, could cause longer time degradation. 
The long-term stability can be monitored with an accuracy of a few percent 
for all the OEV counters with three-hour data taking.

In conclusion, the energy calibration method described in this section 
satisfies the purpose of monitoring light yield variation with sufficient accuracy 
during the data taking of the KOTO experiment.

\section{Summary}

We built and operated the OEV counter subsystem as a part of the KOTO experiment at J-PARC
with an extra-photon-detection capability for 1-MeV deposited energy. 
The subsystem consists of 44 lead-scintillator sandwich type counters (64 modules) 
with different cross sectional shapes. 
They were installed between the CsI crystals and the support cylinder in the endcap. 

We measured the performance of the counters 
and confirmed they met the requirements for the KOTO experiment.  
Light yield and time resolution were measured for each module with cosmic rays before the installation. 
Mean light yield and standard deviation of 64 modules at the middle of the counter were $20.9\pm2.8$~p.e./MeV. 
The minimum light yield in the whole counter area was 15.4~p.e./MeV, 
which is large enough for 1-MeV energy deposit. 
Position dependence of light yield along the fiber direction was 
measured to be about 5\% for all the counters.  
Time resolution dependence on the light yield was observed to be $6.4/\sqrt{n_{\rm p.e.}}\oplus 0.49$~ns. 
With an energy deposit of 1~MeV, the time resolution was 1.5~ns. 
Time resolution was also measured with the KOTO 125-MHz waveform digitizer with 10-pole low-pass filter.  
Results were consistent with those obtained  
with constant-fraction discriminators and TDCs.

An energy calibration method using cosmic rays during no-beam period was developed. % for the physics experiment. 
Stable light yields were observed for all the counters during the data collection period. %for the physics experiment. 
We confirmed that the calibration method works well for various shapes of the OEV counters and 
different stacking directions with an accuracy of a few percent.

\section*{Acknowledgments}
We gratefully acknowledge the support of the staff at J-PARC for providing excellent experimental conditions. 
The authors would also like to express their gratitude to Atsushi Takashima (NDA) and Masayuki Goto (G-Tech Inc.) for 
their continuous support on technical issues during the design and fabrication of the OEV counters.
Part of this work was financially supported by MEXT KAKENHI Grant Number 18071006.

%\label{}

%% The Appendices part is started with the command \appendix;
%% appendix sections are then done as normal sections
%% \appendix

%% \section{}
%% \label{}

%% References
%%
%% Following citation commands can be used in the body text:
%% Usage of \cite is as follows:
%%   \cite{key}         ==>>  [#]
%%   \cite[chap. 2]{key} ==>> [#, chap. 2]
%%

%% References with bibTeX database:

\bibliographystyle{elsarticle-num}
\bibliography{<your-bib-database>}

%% Authors are advised to submit their bibtex database files. They are
%% requested to list a bibtex style file in the manuscript if they do
%% not want to use elsarticle-num.bst.

%% References without bibTeX database:

\end{document}